\documentclass[12pt]{iopart}

\usepackage{graphicx}
\usepackage{dcolumn}
\usepackage{subfigure}
\usepackage{bm}
\usepackage{braket}             
\usepackage[utf8x]{inputenc}    
\usepackage[T1]{fontenc}
\usepackage{textcomp}
\usepackage{siunitx}
\usepackage[normalem]{ulem}     

\usepackage[sort&compress,square,comma,numbers]{natbib}
\newcommand{\Argum}[1]{\ensuremath{\! \left( #1 \right)}}
\newcommand{\MyPar}[1]{\ensuremath{\left( #1 \right)}}

\newcommand{\nat}{Nature}

\begin{document}


\title{Entanglement loss in molecular quantum-dot qubits due to interaction with the environment}

\author{Enrique P. Blair}

\address{%
Electrical and Computer Engineering Department,
Baylor University, 
Waco, TX, USA
}%

\author{G\'eza T\'oth}
\address{Department of Theoretical Physics,
University of the Basque Country
UPV/EHU, P. O. Box 644, E-48080 Bilbao, Spain}
\address{IKERBASQUE, Basque Foundation for Science,
E-48013 Bilbao, Spain}
\address{Wigner Research Centre for Physics, Hungarian Academy of Sciences, P. O. Box 49, H-1525 Budapest, Hungary}

\author{Craig S. Lent}
\address{
Department of Electrical Engineering,}
\address{Department of Physics,
University of Notre Dame,
Notre Dame, IN 46556, USA
}%
 \eads{lent@nd.edu}

\date{\today}

\begin{abstract}

We study quantum entanglement loss due to environmental interaction in a condensed matter system with a complex geometry relevant to recent proposals for computing with single electrons at the nanoscale. 
We consider a system consisting of two qubits, each realized by an electron in a   double quantum dot, which are initially in an entangled Bell state. The qubits are widely separated and each interacts with its own environment. The environment for each is modeled by surrounding double quantum dots placed at random positions with random orientations. We calculate the unitary evolution of the joint system and environment. The global state remains pure throughout. 
We examine the time dependence of the expectation value of the bipartite Clauser-Horne-Shimony-Holt (CHSH) and 
Brukner-Paunkovi\'c-Rudolph-Vedral
(BPRV) Bell operators and explore
the emergence of correlations consistent with local realism.
Though the details of this transition depend on the specific environmental geometry, we show how the results can be mapped on to a universal behavior with appropriate scaling. We determine
the relevant disentanglement times based on realistic physical parameters for molecular double-dots.


\end{abstract}

\maketitle


\section{\label{sec:introduction}Introduction}

Quantum entanglement plays a central role in quantum information science and quantum optics \cite{Horodecki2009Quantum,Guhne2009Entanglement}. 
Due to  recent technological breakthroughs, it is now possible to create entangled states, for instance, with photons, cold trapped ions, ultracold atoms and solid state systems \cite{Schwemmer2014Experimental, Gao2010Experimental,Leibfried2004Toward,Huang2011Multi-partite,Appel2009Mesoscopic,Sewell2012Magnetic,Gross2010Nonlinear,Lucke2014Detecting,Hosten2016Measurement,Mcconnell2015Entanglement,Haas2014Entangled,Bernien2012Heralded,Steffen2006Measurement}. Even if we successfully entangle two particles, they soon will become disentangled due to  quantum correlations built up with the environment. 
Hence, environmentally induced loss of entanglement has  received much recent attention. 

Because the environment involves many degrees of freedom, most models adopt a phenomenological approach which includes, for example, coherence and energy decay times as model inputs. The dynamics of the entangled particles are then non-unitary 
\cite{Gardiner2004Quantum,Breuer2002Theory,Yu598Sudden,Bodoky2009Modeling}.  Few-particle systems can be more tractable yet still illuminate the behavior of much larger systems. For example, recent experiments have been successful at realizing a coherent system-environment dynamics with few particles, and saw thermal behavior emerge in a sub-system, even when the global system evolution is unitary with a fixed energy \cite{Gring2012Relaxation,Kaufman2016Quantum}.

Here, we consider a specific quantum system embedded in a few-particle environment for which we can calculate the global unitary system+environment dynamics exactly. The model environment is large enough for us to observe the disentangling dynamics of the embedded subsystem and see behavior similar to that expected from a large environment. In particular, we examine qubits realized by double quantum dots (DQD) with an extra electron, where the electron position encodes the quantum information.  We study a {\it system} of two DQDs which are initially prepared in a maximally entangled state. These target DQDs are  spatially separate and each interacts Coulombically with its own {\it environment} consisting  of randomly placed similar DQDs with random orientations. Each environmental DQD interacts with the target DQD  and also with all other DQDs in the same environment. Due to this interaction, the target DQDs become disentangled, which we follow by computing the Bell correlation function of  Clauser-Horne-Shimony-Holt (CHSH) \cite{Clauser1969} and also that of Brukner-Paunkovi\'c-Rudolph and Vedral  (BPRV) \cite{Brukner2006Entanglement-Assisted}, and look for the transition to classical (i.e., local realistic) behavior and the time scale on which this happens. Though the system is small enough to calculate the unitary dynamics exactly, it represents a large-dimensional  Hilbert space, so aspects of the behavior of truly large systems emerge. 

The intent of our model is not to propose 
a setup that can readily be
realized in the laboratory.
Rather we explore disentanglement in a concrete physically-motivated system that captures the key elements of the issue---separated but initially entangled pairs, and spatially distinct environments---in a tractable model amenable to direct numerical solution. 

Recent experiments have entangled spatially separated double-dots using photons \cite{GuoCoupledDoubleDots2015, BerndtDoubleDotBell2017}, and phonons \cite{GuoCoupledQCResonator2016}. Remote electron spin systems have been entangled over a distance of more than a kilometer using microwave photons \cite{Hensen2015a} and resulting Bell CHSH violations were measured. We do not concern ourselves here with the details of how the initial entanglement is established, but rather examine its decay due to entanglement with the environment. 

The main characteristics of our model are the following.

(i) The joint state of system and environment remains pure during the dynamics. Thus, we model both the system and environment exactly. There are no stochastic or phenomenological terms added into the model.

(ii) The setup is physically motivated, rather than based on an abstract spin chain model. In fact, this system is a useful model for molecular mixed-valence  double quantum dot systems and has the advantage of including a natural and physically realistic coupling mechanism \cite{Lu2010Charge,Christie2015Synthesis,Quardokus2012Through-Bond,Quardokus2012Adsorption}. Such a double-dot system is a promising candidate for digital computing at the nanoscale \cite{Blair2013}. 

(iii) The two target DQDs do not interact with each other and they have their own environments, which are separated and therefore not coupled to each other.  This  reflects the typical physical situation of spatially separated qubits and avoids artificial environment-mediated entanglement between the target systems.  
 
(iv) There is Coloumbic coupling between the environment DQDs themselves, not only between the system DQDs and the environment DQDs. The coupling strength is computed from the distance and the orientation of the double dots. 

(v) The model is the simplest possible to contain the necessary ingredients. The Hamiltonian consist of terms corresponding to the Coulomb energy only, while the electrons cannot tunnel between the dots. Hence the electrostatic energy remains constant during the dynamics. Moreover, in the computational basis, only the phases of the state vector components change, the amplitude remains constant.

(vi) Remarkably, even if a small number of environment DQDs are considered (we will show results for 10 environmental DQDs below), the decay of entanglement between the system DQDs results in a smooth decay of the Bell correlations. After appropriate normalization, all curves corresponding to various random arrangements of the environment double dots collapse to the same curve.

This model is an extension of a previous studies of the decoherence of a single double dot qubit state due to the environment \cite{Blair2013,Blair2017}. There, entanglement with the environmental drives the local system   into Zurek ``pointer states.'' More complex internal dynamics have also been studied. Mixed valence molecules,  which might realize $1$ nm size double-dot qubits, have additional nuclear motion to consider. Electron transfer from one dot to the other is coupled to vibrational modes of the nuclei, and ultimately the substrate \cite{Blair2016}. In the present model we consider only rigid double-dots to focus on the issue of entanglement loss alone.  

Our paper is organized as follows. In Sec.~2, we describe the model. 
In Sec.~3, we discuss which observables we need to measure to obtain Bell inequality violations. In Sec.~4, we present the results of our calculations in modeling the quantum dynamics of the system in time as quantum entanglement vanishes. 




\begin{figure}[b]
\centering
\includegraphics[width=3.75in]{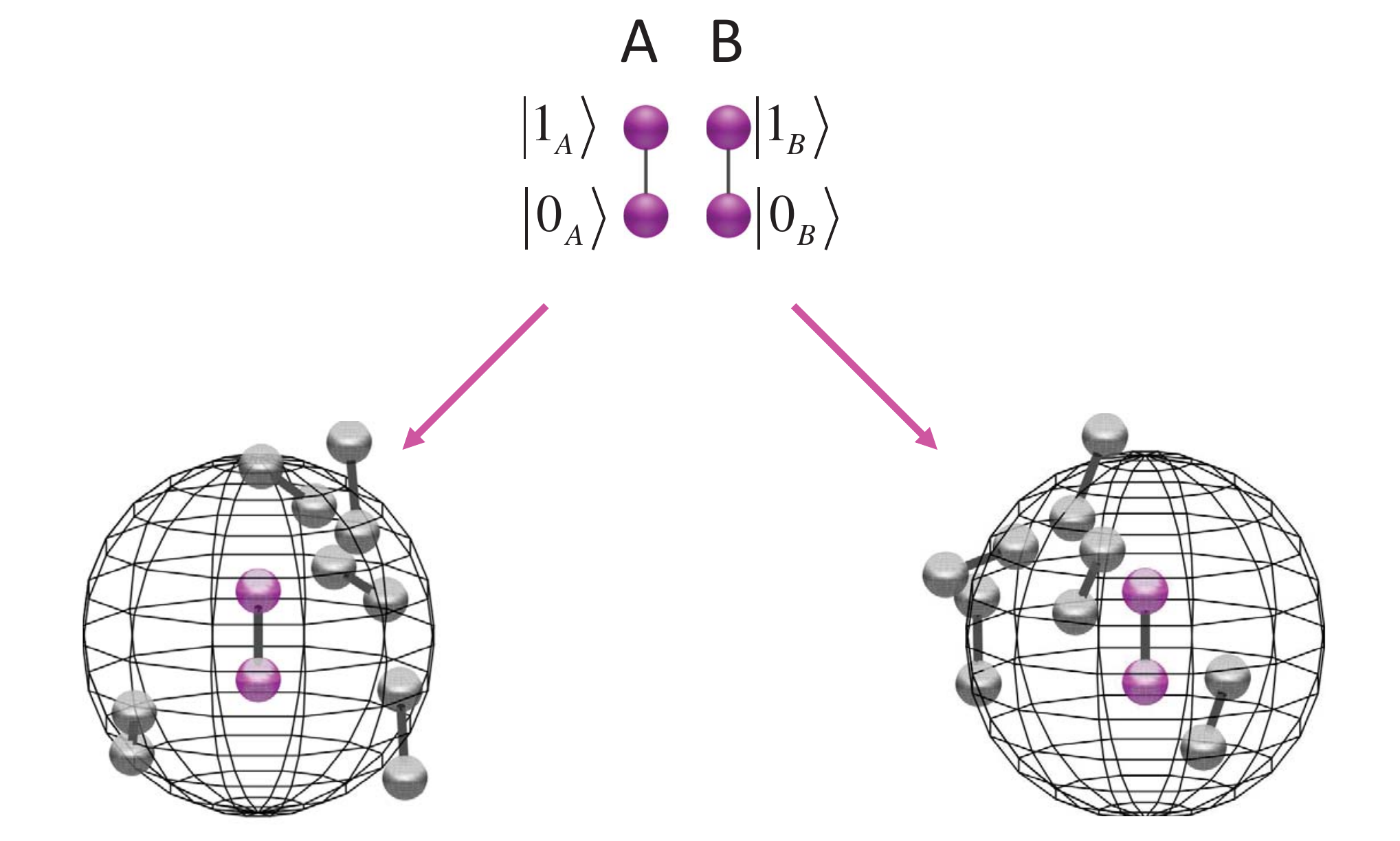}
\caption{A pair of entangled double quantum dots are spatially separate and interact with local environments.  Each pair of spheres represents a double dot with dot separation $a$. The basis states for each pair correspond to a $1$ (top dot occupied)  or $0$ (bottom dot occupied).  The target pair of double-dots (colored in purple) are prepared in an entangled symmetric Bell state. Each interacts with a local environment of similar double dots, randomly positioned and oriented around them in a sphere of radius $R$. The Coulomb interaction couples the double-dot states and the global system evolves under coherent unitary evolution.}
\label{fig:PhysicalPicture}
\end{figure}

\section{\label{sec:ModelDescription} Model description}


The target DQDs labeled $A$ and $B$ are described using a two-state basis composed of states $\Ket{\alpha^A_0}=\Ket{0_A}$, $\Ket{\alpha^A_1}=\Ket{1_A}$, $\Ket{\alpha^B_0}=\Ket{0_B}$, and $\Ket{\alpha^B_1}=\Ket{1_B}$  in which the the electron is fully localized on the bottom ($0$) or top ($1$) dot, respectively. A fixed charge of $+e/2$ resides at each dot, providing net charge neutrality for each DQD. The two initially entangled DQDs are far apart and each interact with a separate environment.

The environment is composed of several similar DQDs surrounding the target DQDs. The environmental DQDs are arranged in a sphere of radius $R$ with the positions on the sphere and the orientation of the double-dots chosen randomly as shown in Fig.\ \ref{fig:PhysicalPicture}. Each of the environmental DQDs, indexed by $k=1,2, \dots N$, is described by a similar two-state basis $\Ket{\alpha^k_0}=\Ket{0_k}$ and $\Ket{\alpha^k_1}=\Ket{1_k}$. There are $N/2$ environmental DQDs surrounding each target DQD ($N$ is always chosen to be even). We label individual DQD basis states with an integer $m\in \{0, 1\}$. Target DQD basis states are indexed with $m_A$ and $m_B$, while the $k_{th}$ environmental DQD is indexed with $m_k$. 

The electronic configuration associated with a specific environmental basis state can then be referred to using the vector 
\begin{equation}
\vec{m}\equiv [m_1, m_2, \dots, m_N].
\end{equation} 
There are $N_E=2^N$ such vectors, 
$[\vec{m}_1, \vec{m}_2, \vec{m}_3, \dots \vec{m}_{N_E}]$, each representing a specific electronic configuration of the environment $E$.

The basis states for describing the global system $\Omega$, which includes the two target DQDs and the environment, consist of the direct products of  the individual DQD states  
\begin{eqnarray}
\Ket{\Phi_{m_A,m_B,\vec{m}_p}}=
               \quad \Ket{ \alpha^A_{m_A}    }
                    \Ket{ \alpha^B_{m_B}    } 
                    \Ket{ \alpha^1_{[\vec{m}_p]_1} }
                    \Ket{ \alpha^2_{[\vec{m}_p]_2} } \dots
                    \Ket{ \alpha^N_{[\vec{m}_p]_N} }.
\label{eq:BasisPhis}
\end{eqnarray}
The global state can then be written as a linear combination of these basis states
\begin{equation}
\Ket{\psi(t)}=
\sum_{
\footnotesize{\begin{array}{c}
m_A,m_B=0,1\\
p=1,2, \dots N_E
\end{array}}
} 
                 c_{m_A,m_B,\vec{m}_p}(t)  \Ket{ \Phi_{m_A, m_B, \vec{m}_p } }.
\label{eq:TotalPsi}
\end{equation}

\subsection{\label{subsec:Hamiltonian} System Hamiltonian}
The Hamiltonian for the global system, including $A$, $B$, and the environment $E$, is determined only by the electrostatic interaction between DQDs in the basis state electronic configurations.  
Let $U^{j,k}_{m_j, m_k}$ be the electrostatic potential energy between the $j^{\mbox{\tiny{th}}}$ qubit 
in state $m_j$ (0 or 1) and the  $k^{\mbox{\tiny{th}}}$ qubit 
in state $m_k$ (0 or 1). This energy is given by
\begin{equation}
U^{j,k}_{m_j, m_k} = \frac{P(m_j) P(m_k) e^2}{16 \pi \epsilon_o} \left[ \frac{1}{r^{j,k}_{0,0}} - \frac{1}{r^{j,k}_{0,1}}  - \frac{1}{r^{j,k}_{1,0}} + \frac{1}{r^{j,k}_{1,1}} \right].
\end{equation}
where $e$ is the fundamental charge, $\epsilon_o$ is the permittivity of free space, $r^{j,k}_{m_j, m_k}$ is the distance between dot $m_j$ in DQD $j$ and dot $m_k$ in DQD $k$, and $P(m)$ is the polarization of a DQD in state $m$. $P(1)=+1$ and $P(0)=-1$. 

The total electrostatic potential energy of a configuration of target DQDs in states $m_A$ and $m_B$, and the environment in the state defined by $\vec{m}_p$ is calculated by simply summing over all interactions between all pairs of DQDs.  
\begin{equation}
E_{m_A,m_B,\vec{m}_p} = \frac{1}{2}\sum_{
                                          j\ne k}                               
                         U^{j,k}_{m_i,m_j}  \; .
\end{equation}
\noindent Here the sums over indices $i$ and $j$ are over the DQDs $[A, B, 1, 2, \dots, N]$, that is, including both target and environmental DQDs.

In this model there is no tunneling between dots within the DQD; we are interested in the entanglement of the phase degrees of freedom rather than electron transfer effects which have been studied elsewhere \cite{Blair2016}.
The Hamiltonian for the global system is then diagonal in the basis states defined by Eq. (\ref{eq:BasisPhis}) is:

\begin{equation}
\hat{H}=\sum_{m_A, m_B, p}
                    \Ket{\Phi_{m_A,m_B,\vec{m}_p}}
                    E_{m_A,m_B,\vec{m}_p}
                    \Bra{\Phi_{m_A,m_B,\vec{m}_p}}.
\label{eq:Htotal}
\end{equation}

We can characterize the strength of the interaction between each target DQD and its local environment by the electrostatic energy needed to flip DQD $A$ or $B$ from $0$ to $1$ with the environment in state $\vec{m}_p$ as
\begin{eqnarray}
E^{\mbox{\tiny{flip}}}_{A,\vec{m}_p}&\equiv&  E_{1,m_B,\vec{m}_p}-E_{0,m_B,\vec{m}_p},\nonumber\\
E^{\mbox{\tiny{flip}}}_{B,\vec{m}_p}&\equiv&  E_{m_a,1,\vec{m}_p}-E_{m_A,0,\vec{m}_p}.
\label{eq:FlipDef}
\end{eqnarray}
These energies depend on the electrostatic configurations of the environment DQDs. We now define a quantity
independent of the quantum state of the environment. Let the root-mean-square  of the flip energies over all the electronic configurations of the local environmental basis states be $E^{\mbox{\tiny{flip}}}_{\mbox{\tiny{RMS}}}(A/B)$. A characteristic time can then be defined for each of the separated systems and the system as a whole as
\begin{eqnarray}
\tau_{A,E} = h/E^{\mbox{\tiny{flip}}}_{\mbox{\tiny{RMS}}}(A), \quad \tau_{B,E} = h/E^{\mbox{\tiny{flip}}}_{\mbox{\tiny{RMS}}}(B), \quad
\tau_E=\sqrt{\tau_{A,E} \tau_{A,E}}.
\label{eq:TauDef}
\end{eqnarray}
The characteristic time $\tau_E$ depends on the details of the geometrically random orientation and positions of the environmental DQDs. As we will see later, while the system is described by a complicated interaction of randomly placed double dots, the time constants given in Eq.~(\ref{eq:TauDef}) characterize the main aspects of the dynamics \cite{Blair2017}.


\subsection{\label{subsec:EquationOfMotion} Density operator and equation of motion}

The time evolution of the system is calculated using the equation of motion of the density operator. 
The density operator for the global system is defined from (\ref{eq:TotalPsi}) by
\begin{eqnarray}
\fl \hat{\rho}_\Omega(t)=\Ket{\psi(t)}\Bra{\psi(t)}\, = \sum_{\footnotesize{\begin{array}{c}
                              m_A, m'_A\\
                              m_B, m'_B\\
                              p, p'\end{array}}
                              }   
{ c_{m'_A,m'_B,\vec{m}_p'}
  c^*_{m_A,m_B,\vec{m}_p } 
  \Ket{\Phi_{m_A,m_B,\vec{m}_p      }}              
 \Bra{\Phi_{m'_A,m'_B,\vec{m}_{p'} }} }.
  \label{eq:RhoDef}
\end{eqnarray}
The dynamics of the global system density matrix is obtained by solving the von Neumann equation as 
\begin{equation}
\hat{\rho}_\Omega \Argum{t} = e^{-i \frac{\hat{H}}{\hbar} t} \, \hat{\rho}_\Omega (0) \,
                       e^{+i \frac{\hat{H}}{\hbar} t} .
\label{eq:Liouville}
\end{equation}
This time evolution is exact within the model and the global system described by $\hat{\rho}_\Omega$ is always in a pure state.



We now define the initial state of the joint system.
The target DQDs $A$ and $B$ are initially in the  symmetric entangled Bell state
\begin{equation}
\Ket{\psi^{AB}(0)}=\left[ \Ket{0}_A\Ket{0}_B+\Ket{1}_A\Ket{1}_B
\right]/\sqrt{2}.
\label{eq:BellState}
\end{equation}
The initial state of the $k_{th}$ environmental DQD is an unpolarized state given as
\begin{equation}
    \Ket{\psi_k \Argum{0}} = e^{i\theta_k} \MyPar{\Ket{0_k} + e^{i \phi_k} \Ket{1_k}}/\sqrt{2},\label{eq:initial_env}
\end{equation}
where the phases  $\theta_k$ and $\phi_k$ are chosen  randomly,  with a distribution that results in the corresponding Bloch vectors being uniformly distributed over the unit sphere. 
We take the initial state of the density operator to be a tensor product state of the entangled system $AB$ and the complete environment. 






\section{\label{sec:measuring_entanglement} Tracking entanglement with Bell operators}

We will primarily observe the disentanglement of the target DQDs by computing the dynamics of the expectation values of  Bell operators. These  are are relevant experimentally, since they can obtained by projective measurements on the subsystems. In the next section we will also calculate entanglement measures for the evolving system.

\subsection{\label{sec:CHSH }CHSH correlation function}

We now calculate the Clauser-Horne-Shimony-Holt (CHSH) correlation function and the corresponding Bell inequality \cite{Clauser1969}. This function is often measured experimentally and it has been shown that states violating the CHSH inequality can be used in the Ekert protocol for entanglement assisted quantum communication \cite{Ekert1991Quantum}.

For each subsystem $A$ and $B$ we define operators in the space spanned by the local basis vectors $\Ket{0}$ and $\Ket{1}$. In this basis we define the rotation operator $\hat{R}$ as
\begin{equation}
\hat{R}(\theta)=\cos(\theta)\left[\, \Ket{1}\Bra{1} + \Ket{0}\Bra{0}\,\right]+
                \sin(\theta)\left[\, \Ket{0}\Bra{1} - \Ket{1}\Bra{0}\,\right].
\end{equation}
We define two basis sets, 
$\bm{a}$ and $\bm{a}'$, for measurements on subsystem $A$  as
\begin{eqnarray}
\Ket{a_l}&=& \hat{R}(\theta_a) \Ket{l}_A,  \nonumber\\
\Ket{a'_l}&=& \hat{R}(\theta_{a'}) \Ket{l}_A
\end{eqnarray}
for $l=0,1$ for indicating the two basis states.
We also define two basis sets, $\bm{b}$ and $\bm{b}'$,  for measurements on subsystem $B$ 
 \begin{eqnarray}     
\Ket{b_l}&=& \hat{R}(\theta_b) \Ket{l}_B, 
\nonumber\\
\Ket{b'_l}&=& \hat{R}(\theta_{b'}) \Ket{l}_B,      
\end{eqnarray}
for $l=0,1.$
For the maximum Bell violation we choose 
$[\theta_a, \theta_{a'}, \theta_b, \theta_{b'}]=[0\si{\degree}, 45\si{\degree}, 22.5\si{\degree},67.5\si{\degree}]$.
 We define  projection operators for measuring the four combinations of $0$ and $1$ on the two parties, for measurements using the $\bm{a}$ and $\bm{b}$ bases as
\begin{equation}
\hat{P}_{kl }(a,b)= \Ket{a_k}\Bra{a_k} \otimes \Ket{b_l}\Bra{b_l},
\label{eq:PplusminusDef}
\end{equation}
for $k,l=0,1.$ The CHSH correlation function encodes the $\Ket{0}$ and $\Ket{1}$ states with a $-1$ and $+1$ respectively.
The expectation value for the product of the measurements ($\pm1$) on $A$ and $B$ using these bases is then given by
\begin{eqnarray}
\hat{P}_{\times}(a,b)&\equiv& P_{00}(a,b) - P_{01}(a,b)-P_{10}(a,b) + P_{11}(a,b),   \nonumber\\
E(a,b)&=&\left< P_{\times} \right> =\Tr\left( \hat{\rho} \hat{P}_{\times}(a,b) \right).
\label{eq:EchshDef}
\end{eqnarray}
Expressions analogous to Eq.(\ref{eq:PplusminusDef}) and (\ref{eq:EchshDef}) define similar quantities $E(a,b')$,  $E(a',b)$, and $E(a',b')$ using the other choices of basis states. The CHSH correlation function is then defined to be
\begin{equation}
S_{\mbox{\tiny{CHSH}}}=\left| E(a,b) -E(a,b') + E(a',b) + E(a',b') \right|.
\label{eq:CHSHcorrelation}
\end{equation}
The assumption of local realism yields the Bell inequality 
\begin{equation} 
S_{\mbox{\tiny{CHSH}}} \leq 2.
\label{eq:CHSHinequality}
\end{equation}
\noindent For local  values of $a, a',b,$ and $b'$ which are distributed randomly and uniformly $S_{\mbox{\tiny{CHSH}}} = \sqrt{2}$ holds, 
obeying the inequality. 
For the fully entangled Bell state of Eq. (\ref{eq:BellState}), by contrast,  
$ S_{\mbox{\tiny{CHSH}}} = 2\sqrt{2}$, in violation of  (\ref{eq:CHSHinequality}).

\subsection{\label{sec:MerminBell}Brukner-Paunkovi\'c-Rudolph-Vedral  correlation function}

We will now consider the Bell inequality derived by \v C. Brukner {\it et al.,}
in Ref.~\cite{Brukner2006Entanglement-Assisted}, which is a generalization of Mermin's Bell inequality \cite{Mermin1985,Maccone2013}
for bipartite states that are not necessarily symmetric. 

We define three sets of rotated basis function as
\begin{equation}
\Ket{u_k}= \hat{R}(\theta_k) \Ket{0}, \quad\quad \Ket{v_k}= \hat{R}(\theta_k) \Ket{1},
\end{equation}
for $k=1,2,3.$
For the correlation function we choose 
$[\theta_1, \theta_2, \theta_3]=[0\si{\degree}, 120\si{\degree}, 240\si{\degree}]$. 
For each subsystem $A$ and $B$ a particular basis set corresponding to one of these angles is randomly chosen and a projective measurement is carried out.
The projection operators corresponding to measurements are
\begin{equation}
\hat{P}_k^{(0)} = \Ket{u_k}\Bra{u_k}, \quad\quad \hat{P}_k^{(1)} = \Ket{v_k}\Bra{v_k},
\end{equation}
where we used the subscript $k=1,2,3$ for the three measurement settings.
Each of these operators has eigenvalues $0$ and $1$, hence all measurements on either subsystem $A$ or $B$ have these outcomes. 
We define the correlation operators for obtaining the same outcome in the two qubits as
\begin{equation}
\hat{P}_{kl}^{\rm same}(A,B)=\hat{P}_k^{(0)}(A) \otimes \hat{P}_l^{(0)}(B)
 + \hat{P}_k^{(1)}(A) \otimes \hat{P}_l^{(1)}(B),  
\label{eq:ProjectionDefs}
\end{equation}
and  for obtaining the opposite outcomes as
\begin{equation}
\hat{P}_{kl}^{\rm opp}(A,B)=\hat{P}_k^{(0)}(A) \otimes \hat{P}_l^{(1)}(B)
 + \hat{P}_k^{(1)}(A) \otimes \hat{P}_l^{(0)}(B),  
\label{eq:ProjectionDefs2}
\end{equation}
for $k\ne l.$
The operator  $\hat{P}_{12}^{\rm same}(A,B)$, for example,  corresponds to a measurement of setting $1$ on one subsystem and setting $2$ on the other, both yielding the same result (both $0$ or both $1$).

The correlation function needed to evaluate the Bell inequality is then defined as 
\begin{eqnarray}
S_{\mbox{\tiny{BPRV}}}= \sum^3_{k}\left \langle \hat{P}_{kk}^{\rm same}(A,B) \right\rangle + 
\sum^3_{k,l\ne k} \left \langle \hat{P}_{kl}^{\rm opp}(A,B) \right\rangle ,
\label{eq:MerminCorrelation}
\end{eqnarray}
where the subscript refers to the initials of the authors of Ref.~\cite{Brukner2006Entanglement-Assisted}.
Local realism requires that  each subsystem $A$ and $B$ have values that determine the results of measurements of settings $1$, $2$, and $3$ before the measurement is made.  That assumption yields the inequality
\begin{equation}
S_{\mbox{\tiny{BPRV}}}\leq 7
\label{eq:MerminBellInequality}
\end{equation}
for any probability distribution of the measurement outcomes \cite{Maccone2013}.
By contrast, the fully entangled Bell state of Eq. (\ref{eq:BellState}) yields $S_{\mbox{\tiny{BPRV}}}=7.5$ in clear violation of Eq. (\ref{eq:MerminBellInequality}). 
We evaluate directly the time-dependent value $S_{\mbox{\tiny{BPRV}}}(t)$ from the density matrix evolving in time under Eq. (\ref{eq:Liouville}).

\section{Results}


Starting with the initial state given by Eqs.~(\ref{eq:BellState}) and (\ref{eq:initial_env}), we solve for the unitary evolution of the global density matrix using Eq.~(\ref{eq:Liouville}), and calculate the correlation functions  $ S_{\mbox{\tiny{CHSH}}}(t)$  and $ S_{\mbox{\tiny{BPRV}}}(t)$ directly from the global density matrix.  The number of environmental double-dots is  $N_E=10$ (five around each target double-dot) yielding $2^{10}=1024$ environmental electronic configurations. 

\begin{figure}[htbp] 
   \includegraphics[width=6in]{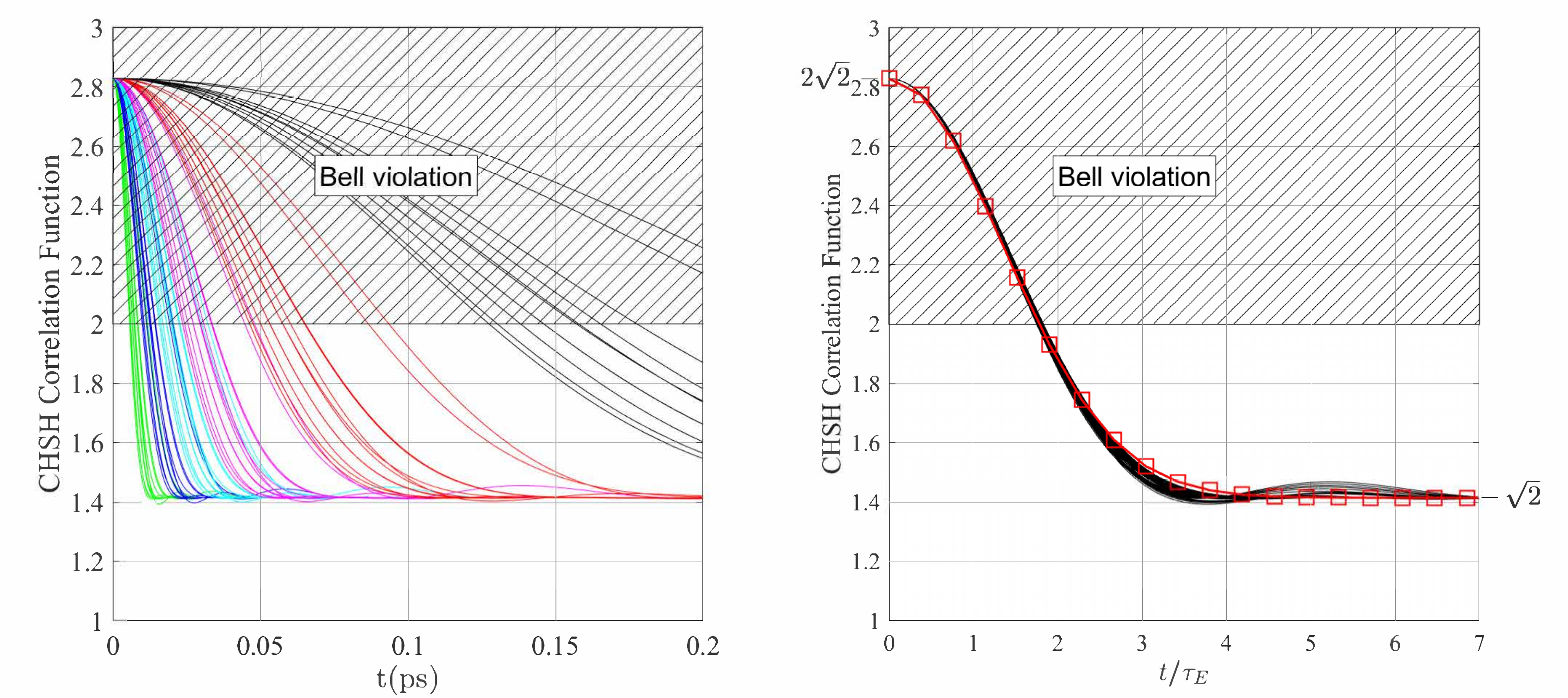} 
   \caption{The dynamics of the Bell correlation functions indicate the loss of entanglement  under unitary evolution. 
    (a) The  CHSH correlation function $ S_{\mbox{\tiny{CHSH}}}(t)$, defined by equation (\ref{eq:CHSHcorrelation}), is calculated as functions of time, assuming unitary evolution of the global system. The curves are shown for $a=1$~nm (which sets the time-scale) and for  6 different values of $R/a= \{2.5, 3, 3.5, 4, 5, 7\}$, corresponding to different average strengths of interaction with the environment. For increasing values of $R/a$ the lines are colored [green,  blue, cyan, magenta, red, black]. For each value of $R/a$, the results for 12 different random geometrical configurations of the environment are shown. 
     (b) Scaled dynamics of the CHSH correlation functions for different random geometries of the environment.  The value of the CHSH correlation function for all 72 different geometrical configurations of the environment shown in (a) are  plotted here versus the time scaled by the characteristic time $\tau_E$ calculated from Eq.~(\ref{eq:TauDef}).  The time $\tau_E$ depends on the mean energy of interaction with the environment and  is different for each random geometry. The value of the CHSH correlation function decays from $2\sqrt{2}$ for the pure  Bell state to $\sqrt{2}$ for the classical mixture.
   The points (squares) show the result of a Gaussian fit with characteristic time $\tau_{\mbox{\tiny{opt}}}=1.34$~$\tau_E$.}
   \label{fig:CHSHBellViolation}
\end{figure}

In Fig.~\ref{fig:CHSHBellViolation}(a), the results are shown for $a=1$ nm, a typical scale for molecular double-dots, which sets the time scale at picoseconds. Calculated Bell correlation functions are shown for 6 different values of $R/a\in \{2.5, 3, 3.5, 4, 5, 7\}$, corresponding to different average strengths of coupling to the environment. For each value of $R/a$, 12 different random geometric arrangements of the environments are shown. 

The CHSH correlation function $S_{\mbox{\tiny{CHSH}}}(t)$ for each of these 72 configurations shown on the upper part of Fig.~\ref{fig:CHSHBellViolation},  begins at the value corresponding to maximally violating the  Bell inequality. Then, it starts dropping out of the Bell violation regime and decaying to the classical limit. As expected, the stronger the coupling to the environment, the faster the quantum entanglement disappears.

Fig.~\ref{fig:CHSHBellViolation}(b) shows the same 72 cases of the geometrically random environment as Fig.~\ref{fig:CHSHBellViolation}(a), but plotted on a time axis scaled by the characteristic time  $\tau_E$ as calculated from Eq.~\ref{eq:TauDef}. The value of $\tau_E$ is distinct for each of the random geometries of the environment. The time-scaled result is independent of the values of $a$ or $R/a$. 

The squares show the value of a single Gaussian fit to all 72 curves for the transition from the initial Bell-state value to the classical limit. The fit yields a Gaussian width of $\tau_{\mbox{\tiny{opt}}}=1.34$~$\tau_E$ and matches the calculated bundle of trajectories well. The CHSH correlation function evolves from the fully entangled value to the value corresponding to local realism  over a time on the scale of  $\tau_E$ and the transition is very close to Gaussian, rather than the often-assumed exponential associated with semigroup behavior. Note that the slope at small times is here zero, in contrast to the finite slope of an exponential.  

\begin{figure}[htbp] 
   \centering
   \includegraphics[width=5in]{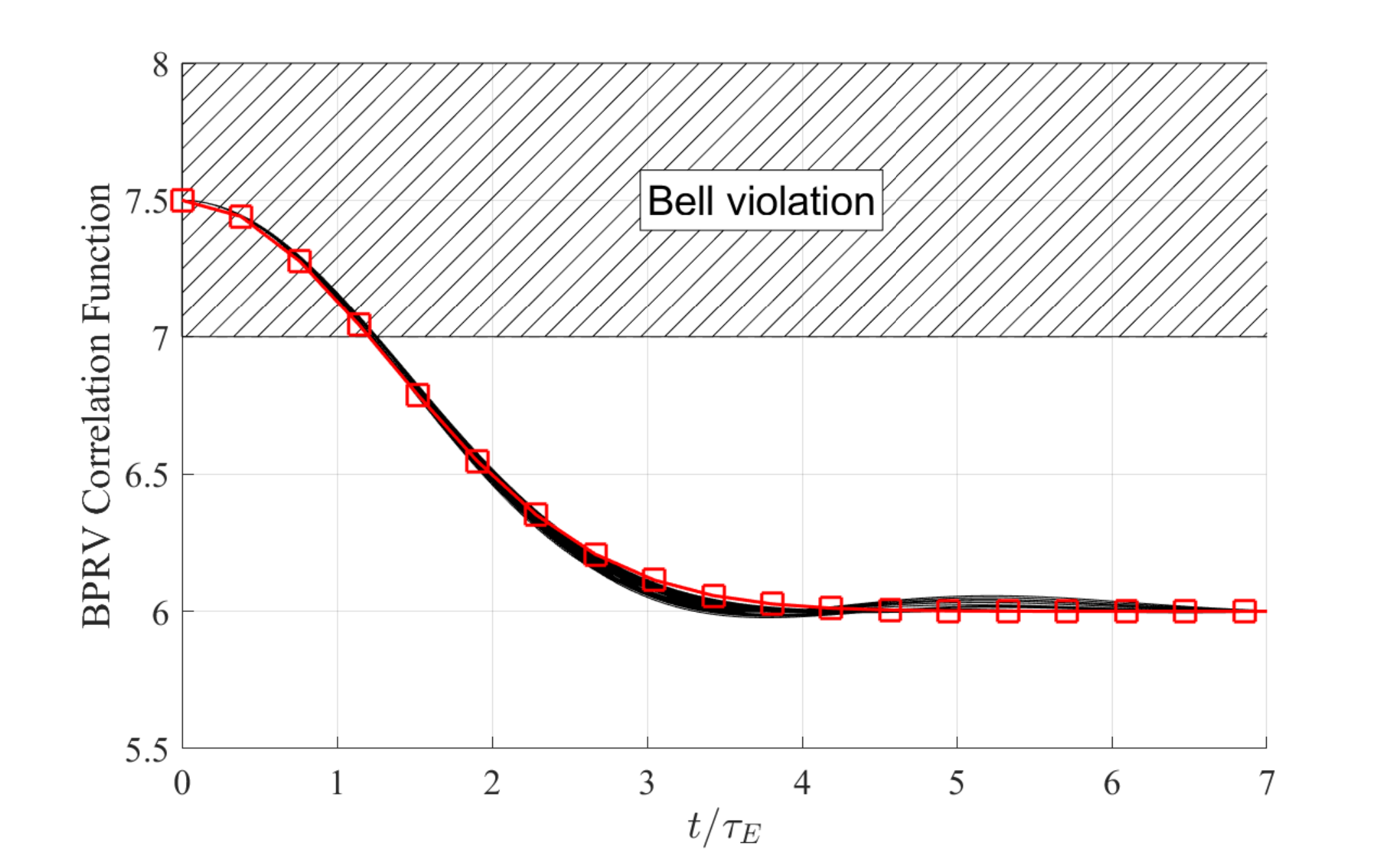}
   
   (a)
  \vskip0.5cm
   
   \includegraphics[width=5in]{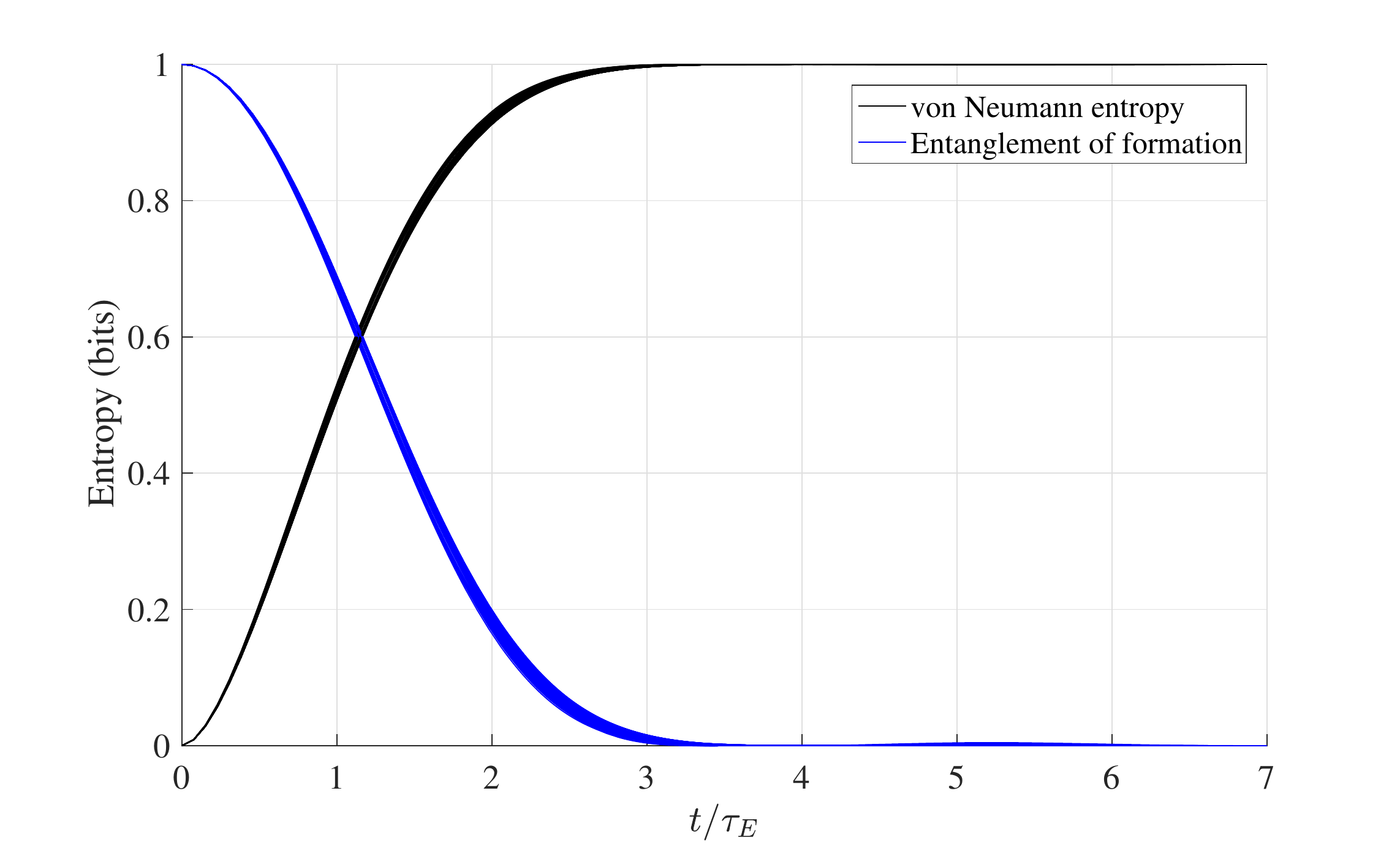} 
   
   (b)
   \caption{(a) Scaled dynamics of the BPRV correlation function (Eq. (\ref{eq:MerminCorrelation})) for 72 different random geometries of the environment. (b) Dynamics of the entanglement of formation of the two system qubits and the von Neumann entropy of their reduced state.  For details, see caption of Fig.~\ref{fig:CHSHBellViolation}(b).
   }
   \label{fig:BPRVBellViolation}
\end{figure}

Fig.~\ref{fig:BPRVBellViolation}(a)  shows the  time-scaled BPRV correlation, which behaves similarly, crossing out of the Bell violation regime and  into the classical (local realism) limit with a Gaussian form. The width of the Gaussian is identical to that for the CHSH correlation function; the squares on the plot show the Gaussian fit. It may be that as the number of environmental DQDs increases, the phase interference that results in the slight residual oscillations averages out to a precisely  Gaussian shape.  

Though both correlation functions have the same Gaussian shape and the same width, the transition out of the region forbidden by local realism occurs at different times for the two different Bell correlation functions. For the BPRV correlation function, the transition occurs at $t/\tau_E\approx 1.21$, whereas for the CHSH correlation function it occurs at $t/\tau_E\approx 1.78$. Of course entanglement can persist even after the system no longer violates a particular Bell inequality. 

The Gaussian shape of the transition into local realism is notable. A similar Gaussian characteristic has been observed by Cucchietti {\em et al.} \cite{Cucchietti2005} in the context of the decoherence of a model spin system.   This result seems to hold across many distributions of coupling to the environment and is  rooted in the approximately Gaussian distribution of the eigenvalues of the Hamiltonian for a random environment.  


Fig.~\ref{fig:BPRVBellViolation}(b) shows the dynamics of the von Neumann entropy of the reduced density operator for the two target qubits, $\rho^{AB}(t)=\Tr_{E}(\rho_\Omega(t))$. The increase in entropy by one bit corresponds to the loss of local information about the state of the pair. The global von Neumann entropy remains zero throughout because the global state is always pure. 

Fig.~\ref{fig:BPRVBellViolation} also shows the time dependence of the entanglement  of formation for the two target qubits \cite{Wootters1998Entanglement}. This quantity decreases as the entanglement between the two DQDs smoothly vanishes.


\section{Conclusion}


We have examined the dynamics of qubits realized with a double-dot with an extra electron. Initially, they were placed in a maximally entangled symmetric Bell state. Then, each interacted Coulombically  with its own environment consisting of similar double-dots with a random position and orientation. While the dynamics was unitary and the entire system remains pure, the two system double-dots become gradually disentangled from each other. We followed this by calculating the dynamics of the expectation values of Bell operators for relevant Bell inequalities. We used the physical parameters of molecular mixed-valence double quantum dot systems for modeling, and calculated the relevant times scales. While the dynamics were different for different random geometries, with appropriate normalization all curves collapsed to a single curve. The time scale of disentanglement 
could be calculated precisely from the mean interaction strength between the system and  the environmental degrees of freedom. 

Our results can be understood by noting that quantum entanglement is best characterized not as fragile, but rather as promiscuous. Dynamics entangles each system with all the other systems with which it interacts. This promiscuity is constrained by the principle of the quantum monogamy of entanglement \cite{Coffman2000, Osborne2006}, which bounds the strength of entanglement between any two pairs when a system entangles with many other systems. The decay of entanglement we see  between the two target DQD systems occurs precisely because they each entangle with multiple systems in their respective environments, while maintaining global coherence and purity completely. 




\ack
We acknowledge support from the National Science Foundation (Grant No. DGE-1313583). We also acknowledge the support of the  EU (ERC Starting Grant 258647/GEDENTQOPT, COST Action CA15220, QuantERA CEBBEC), the Spanish Ministry of Economy, Industry and Competitiveness and the European Regional Development Fund FEDER through Grant No. FIS2015-67161-P (MINECO/FEDER, EU), the Basque Government (Grant No. IT986-16), and the National Research, Development and Innovation Office NKFIH (Grant No.  K124351).


\providecommand{\newblock}{}

\end{document}